\documentclass[twocolumn,showpacs,preprintnumbers,amsmath,amssymb,epsfig]{revtex4}

\usepackage{graphicx}
\usepackage{dcolumn}
\usepackage{bm}
\input{epsf}

\def\apj #1 #2 #3 {#1, ApJ, {\bf #2}, #3}
\def\apjl #1 #2 #3 {#1, ApJ, {\bf #2}, L#3}
\def\apjs #1 #2 #3 {#1, ApJS, {\bf #2}, #3}
\def\aap  #1 #2 #3 {#1, A\&A, {\bf #2}, #3}
\def\mnras #1 #2 #3 {#1, MNRAS, {\bf #2}, #3}
\def\pra #1 #2 #3 {#1, Phys.~Rev.~A., {\bf #2}, #3}
\def\prb #1 #2 #3 {#1, Phys.~Rev.~B., {\bf #2}, #3}
\def\prc #1 #2 #3 {#1, Phys.~Rev.~C., {\bf #2}, #3}
\def\prd #1 #2 #3 {#1, Phys.~Rev.~D., {\bf #2}, #3}
\def\pre #1 #2 #3 {#1, Phys.~Rev.~E., {\bf #2}, #3}
\def\prl #1 #2 #3 {#1, Phys.~Rev.~Lett., {\bf #2}, #3}
\def\plb #1 #2 #3 {#1, Phys.~Lett.~B., {\bf #2}, #3}
\def\science #1 #2 #3 {#1, Science., {\bf #2}, #3}
\def\nature #1 #2 #3 {#1, Nature., {\bf #2}, #3}
\def\nphysa #1 #2 #3 {#1, Nucl.~Phys.~A., {\bf #2}, #3}
\def\nphysb #1 #2 #3 {#1, Nucl.~Phys.~B., {\bf #2}, #3}
\def\nphysbs #1 #2 #3 {#1, Nucl.~Phys.~B.~Suppl., {\bf #2}, #3}

\def\h#1{\hbox{${}^{#1}$H}}

\def\h502{\hbox{$ h^{2}_{50}$}}

\def\la{\mathrel{\mathpalette\fun <}}

\def\fun#1#2{\lower3.6pt\vbox{\baselineskip0pt\lineskip.9pt
  \ialign{$\mathsurround=0pt#1\hfil##\hfil$\crcr#2\crcr\sim\crcr}}}

\newcommand{\vct}[1]{\mbox{\boldmath${#1}$}}

\begin{document}

\title{%
Cosmological Signatures of the Interaction \\
between Dark-Energy and Massive Neutrinos
}%

\author{Kiyotomo Ichiki$^{1,3}$
\footnote{Email address: ichiki@resceu.s.u-tokyo.ac.jp,}
Yong-Yeon Keum$^{2,3}$
\footnote{Email address: yykeum@phys.ntu.edu.tw (Corresponding Author). } ,
}
\affiliation{%
$^1$Research Center for the Early Universe, University of Tokyo, Bunkyo-ku, Tokyo 113-0033, Japan }%

\affiliation{%
$^2$Department of Physics, National Taiwan University, 
Taipei, Taiwan 10672, R.O.C.}%

\affiliation{%
$^3$Theory Division, National Astronomical Observatory, Mitaka, Tokyo 
181-8588, Japan
}%


\vskip1.0cm
\begin{abstract}
We investigate whether interaction between massive neutrinos and quintessence
scalar field is the origin of the late time accelerated expansion of the universe.
We present cosmological perturbation theory in neutrinos probe
interacting dark-energy models, and calculate cosmic microwave
background anisotropies and matter power spectrum. In these models,
the evolution of the mass of neutrinos is determined by the
quintessence scalar field, which is responsible for the cosmic
acceleration today. We consider several types of scalar field
potentials and put constraints on the coupling parameter between
neutrinos and dark energy. Assuming the flatness of the universe,
the constraint we can derive from the current observation is $\sum
m_{\nu} < 0.87 eV$ at the 95 $\%$ confidence level for the sum over
three species of neutrinos. We also discuss on the stability issue of 
the our model and on the impact of the scattering term in Boltzmann
equation from the mass-varying neutrinos.
\end{abstract}



\thispagestyle{empty}
\maketitle
\newpage

{\it \bf Interoduction:} 
After SNIa\cite{sn1a} and WMAP\cite{wmap} observations during last
decade, the discovery of the accelerated expansion of the universe
is a major challenge of particle physics and cosmology.
In order to understand  unknown $76\%$ components of the critical density
of the universe with a negative pressure (dark-energy),
the positive cosmological constant term seems to be a serious candidate 
for the dark energy. 
In this case the cosmological constant $\Lambda$ and it's energy density
$\Lambda/8\pi G$ remain constant with time and correspnding mass density
$\rho_{\Lambda} = 6.44 \times 10^{-30} (\Omega_{\Lambda}/0.7)(h/0.7) \,g cm^{-3} 
$, where h is the Hubble constant $H_0$ expressed in units of 100 
$km s^{-1}Mpc^{-1}$ and $\Omega_{\Lambda}=0.76$.
Although cold dark matter model with a positive cosmological constant \cite{lambda} 
($\Lambda CDM$) provides an excellent explanation of the SN1a data,
the present value of $\Lambda$ is $~10^{123}$ times smaller than the value
predicted by the particle physics model. Among many alternative candidates for 
dark energy\cite{quintessence,brane-world,mgrav} 
the scalar field model like quintessence
is a simple model with time dependent $w$, which is generally larger
than $-1$. Because the different $w$ leads to a different expansion
history of the universe, the geometrical measurements of  
cosmic expansion through observations of SNIa, CMB, and Baryon Acoustic
Oscillations (BAO) can give us tight constraints on $w$.
Further, if the dark energy is dynamical component
like a scalar field, it should carry its density
fluctuations. Thus, the probes of density fluctuations near
the present epoch, such as cross correlation studies of the integrated
Sachs-Wolfe effect \cite{Crittenden:1995ak,Hu:2004yd} and the power of
Large Scale Structure (LSS) \cite{Takada:2006xs}, can also provide useful
information to discriminate between cosmological constant and others. 
Yet, current observational data can give only poor constraints on   
the properties of dark energy fluctuations \cite{Hannestad:2005ak,Ichiki:2007vn}.
Another interesting way to study the scalar field dark energy models is
to investigate the coupling between the dark energy and the other matter
fields. In fact, a number of models which realize the interaction
between dark energy and dark matter, or even visible matters, have been
proposed so far
\cite{Carroll:1998zi,Bean:2000zm,Farrar:2003uw,das-khoury:2006,Lee:2006za}. 
Observations of the effects of these interactions will offer an unique
opportunity to detect a cosmological scalar field
\cite{Carroll:1998zi,Liu:2006uh}. 

{\it \bf Interacting Dark-Energy with Neutrinos:}
In this letter we investigate the cosmological
implication of an idea of the dark-energy interacting with neutrinos
\cite{{Fardon:2003eh},{mavanu}}. For simplicity, we consider the
case that dark-energy and neutrinos are coupled such that the mass
of the neutrinos is a function of the scalar field which drives the
late time accelerated expansion of the universe. In previous works
\cite{{Fardon:2003eh},{mavanu}}, potential term was treated as a
dynamical cosmology constant, which can be applicable for the
dynamics near present epoch, but kinetic energy term has been ignored in thier discussions. However the kinetic contributions
become important to describe cosmological perturbations in early
stage of universe, which is fully considered in our work.

Equations for quintessence scalar field are given by
\begin{eqnarray}
\ddot{\phi}&+&2{\cal H}\dot\phi+a^2\frac{d V_{\rm eff}(\phi)}{d\phi}=0~,
 \label{eq:Qddot}\\
V_{\rm eff}(\phi)&=&V(\phi)+V_{\rm I}(\phi)~,\\
V_{\rm I}(\phi)&=&a^{-4}\int\frac{d^3q}{(2\pi)^3}\sqrt{q^2+a^2
 m_\nu^2(\phi)}f(q)~,\\
m_\nu(\phi) &=& \bar m_i e^{\beta\frac{\phi}{M_{\rm pl}}}~,
\end{eqnarray}
where $V(\phi)$ is the potential of quintessence scalar field, $V_{\rm
I}(\phi)$ is additional potential due to the coupling to neutrino
particles \cite{Fardon:2003eh,Bi:2003yr},
and $m_\nu(\phi)$ is the mass of neutrino coupled to the scalar field.
${\cal H}$ is ${\dot a}/{a}$, where the dot represents the
derivative with respect to the conformal time $\tau$.

Here we consider three different types of the quintessence potential:
(1) inverse power law potentials (Model I), 
(2) SUGRA type potential models (Model II),
(3) exponential type potentials (Model III),
which are given, respectively:
\begin{equation}
M^{4}\left({M_{pl}\over \phi}\right)^{\alpha}~ \hspace{1mm}; \hspace{1mm}
M^{4} \left({M_{pl}\over \phi}\right)^{\alpha} e^{3\phi^2/2M_{\rm
pl}^2}~\hspace{1mm}; \hspace{1mm}
M^4 e^{-\alpha ({\phi \over M_{pl}})}.
\end{equation}
Energy densities of mass varying neutrino (MVN) and quintessence scalar
field are described as
\begin{eqnarray}
\rho_\nu &=& a^{-4}\int \frac{d^3 q}{(2\pi)^3} \sqrt{q^2+a^2m_\nu^2} f_0(q)~, \label{eq:rho_nu}\\
3P_\nu &=& a^{-4}\int \frac{d^3 q}{(2\pi)^3} \frac{q^2}{\sqrt{q^2+a^2m_\nu^2}}
 f_0(q)~, \label{eq:P_nu}\\
\rho_\phi &=& \frac{1}{2a^2}\dot\phi^2+V(\phi)~, \hspace{3mm}
P_\phi = \frac{1}{2a^2}\dot\phi^2-V(\phi)~.
\end{eqnarray}
From equations (\ref{eq:rho_nu}) and (\ref{eq:P_nu}), the equation of
motion for the background energy density of neutrinos is given by
\begin{equation}
\dot\rho_{\nu}+3{\cal H}(\rho_\nu+P_\nu)=\frac{\partial \ln
 m_\nu}{\partial \phi}\dot\phi(\rho_\nu -3P_\nu)~.
\end{equation}
The evolution of neutrinos requires solving the Boltzmann equations 
in the case: \cite{ichiki-yyk:2007a, yyk:2007}:
\begin{equation}
 \frac{dq}{d\tau}=-\frac{1}{2}\dot{h_{ij}}qn^in^j-a^2
  \frac{m}{q}\frac{\partial m}{\partial x^i}\frac{dx^i}{d \tau} ~.
\label{eq:eq-b}
\end{equation}
Our analytic formula in eq.(\ref{eq:eq-b}) is different
from those of \cite{Brookfield-b} and \cite{Zhao:2006zf}, since they
have omitted the contribution of the varying neutrino mass term. 
The first order Boltzmann equations written in the synchronous gauge reads \cite{Ma:1995ey}:
\begin{eqnarray}
 \frac{\partial \Psi}{\partial
 \tau}+i\frac{q}{\epsilon}(\vct{\hat{n}}\cdot\vct{k})\Psi+\left(\dot\eta-(\vct{\hat
 k}\cdot\vct{\hat n})^2\frac{\dot h+6\dot\eta}{2}\right)\frac{\partial \ln
 f_0}{\partial \ln q} && \nonumber \\
= -i\frac{q}{\epsilon}(\vct{\hat{n}}\cdot\vct{k})k\delta\phi\frac{a^2
 m^2}{q^2}\frac{\partial \ln m}{\partial \phi}\frac{\partial \ln
 f_0}{\partial \ln q}~. &&
\label{eq:boltzmann}
\end{eqnarray}
The Bolzmann hierarchy for neutrinos, obtained expanding the perturbation $\Psi$
in a Legendre series can be written as \cite{ichiki-yyk:2007a,yyk:2007} :
\begin{eqnarray}
 \dot{\Psi_0}&=&-\frac{q}{\epsilon}k\Psi_1
                +\frac{\dot h}{6}\frac{\partial \ln{f_0}}{\partial\ln{q}}~, \label{eq:dot_Psi_0}\\
\dot{\Psi_1}&=&\frac{1}{3}\frac{q}{\epsilon}k\left(\Psi_0-2\Psi_2\right)
              + \kappa~, \label{eq:dot_Psi_1}\\
\dot{\Psi_2}&=&\frac{1}{5}\frac{q}{\epsilon}k(2\Psi_1-3\Psi_3)-\left(\frac{1}{15}\dot{h}+\frac{2}{5}\dot{\eta}\right)\frac{\partial \ln{f_0}}{\partial \ln{q}}~,\\
\dot{\Psi_\ell}&=&\frac{q}{\epsilon}k\left(\frac{\ell}{2\ell+1}\Psi_{\ell-1}-\frac{\ell+1}{2\ell+1}\Psi_{\ell+1}\right)~.
\end{eqnarray}
where
\begin{equation}
\kappa = -\frac{1}{3}\frac{q}{\epsilon}k\frac{a^2 m^2}{q^2}\delta\phi
 \frac{\partial \ln m_\nu}{\partial \phi}\frac{\partial \ln
 f_0}{\partial \ln q}\label{eq:kappa}~.
\end{equation}
{\it \bf Constrains on the MaVaNs parameters:}
As was discussed in the introduction, the coupling between cosmological
neutrinos and dark energy quintessence could modify the CMB and matter
power spectra significantly.  It is therefore possible and also
important to put constraints on coupling parameters from current observations.
For this purpose, we use the WMAP3 \cite{Hinshaw:2006ia,Page:2006hz} and
2dFGRS \cite{Cole:2005sx} data sets.

The flux power spectrum of the Lyman-$\alpha$ forest can be used to
measure the matter power spectrum at small scales around $z\la 3$ \cite{McDonald:1999dt,Croft:2000hs}.
It has been shown, however, that the resultant constraint on neutrino
mass can vary significantly from $\sum m_\nu < 0.2$eV to $0.4$eV
depending on the specific Lyman-$\alpha$ analysis used \cite{Goobar:2006xz}.
The complication arises because the result suffers from the
systematic uncertainty regarding to the model for the intergalactic
physical effects, i.e., damping wings, ionizing radiation fluctuations,
galactic winds, and so on \cite{McDonald:2004xp}.
Therefore, we conservatively omit the Lyman-$\alpha$ forest data from
our analysis.

Because there are many other cosmological parameters than the MaVaNu
parameters, we follow the Markov Chain Monte Carlo(MCMC) global fit
approach \cite{MCMC} to explore the likelihood space and marginalize
over the nuisance parameters to obtain the constraint on
parameter(s) we are interested in. Our parameter space consists of
\begin{equation}
\vec{P}\equiv (\Omega_bh^2,\Omega_ch^2,H,\tau,A_s,n_s,m_i,\alpha,\beta)~,
\end{equation}
where $\omega_bh^2$ and $\Omega_ch^2$ are the baryon and CDM densities
in units of critical density, $H$ is the hubble parameter, $\tau$ is the
optical depth of Compton scattering to the last scattering surface, $A_s$
and $n_s$ are the amplitude and spectral index of primordial density
fluctuations, and $(m_i,\alpha,\beta)$ are the parameters of MaVaNs. 
\begin{table}[bth]
\caption{Global analysis data within $1\sigma$ deviation for different
types of the quintessence potential.}
{\begin{tabular}{@{}c|c|c|c|c@{}} \toprule
Quantites & Model I
&Model II & Model III & WMAP-3 data \\ \colrule
$\Omega_B\, h^2[10^2]$ & $2.21\pm 0.07$ & $2.22\pm 0.07$ & $2.21\pm 0.07$ & $2.23\pm 0.07$   \\
$\Omega_{CDM}\, h^2[10^2]$ & $11.10\pm 0.62$ & $11.10\pm 0.65$ & $11.10 \pm 0.63$ & $12.8\pm 0.8$ \\
$H_0$ & $65.97 \pm 3.61$ & $65.37\pm 3.41$ & $65.61\pm 3.26$ & $72\pm 8$  \\
$Z_{re}$ & $10.87 \pm 2.58$ & $10.89 \pm 2.62$ & $11.07 \pm 2.44$ &   ---     \\
$\alpha$ & $< 2.63$ & $< 7.78$ & $< 0.92$ & ---       \\
$\beta$ & $< 0.46 $ & $< 0.47$ & $< 0.58$ & ---       \\
$n_s$ & $0.95\pm 0.02$ & $0.95\pm 0.02$ & $0.95\pm 0.02$ & $0.958\pm 0.016$      \\
$A_s[10^{10}]$ & $20.66\pm 1.31$ & $20.69\pm 1.32$ & $20.72\pm 1.24$ & ---- \\
$\Omega_{Q}[10^2]$ & $68.54\pm 4.81$ & $67.90\pm 4.47$ & $68.22\pm 4.17$ &
$71.6\pm 5.5$      \\
$Age/Gyrs$ & $13.95\pm 0.20$ & $13.97\pm 0.19$ & $13.69\pm 0.19$ &  $13.73\pm 0.16$ \\
$\Omega_{MVN}\,h^2[10^2]$ & $< 0.44$ & $< 0.48$ & $< 0.48$ &  $< 1.97 (95\% C.L.)$ \\
$\tau$ & $0.08\pm 0.03$ & $0.08 \pm 0.03$ & $0.09 \pm 0.03$ & $0.089 \pm 0.030$  \\ \botrule
\end{tabular} \label{ta1}}
\end{table}

Larger $\beta$ will generally lead larger $m_\nu$ in
the early universe. This means that the effect of neutrinos on the
density fluctuation of matter becomes larger leading to the larger
damping of the power at small scales. A complication arise because the
mass of neutrinos at the transition from the ultra-relativistic regime
to the non-relativistic one is not a monotonic function of $\beta$. 
Even so, the coupled neutrinos give larger decrement
of small scale power, and therefore one can limit the coupling parameter
from the large scale structure data.

As shown in table \ref{ta1}, 
we find no observational signature which favors the coupling between
MaVaNs and quintessence scalar field, 
but we don't need to finetune the coupling parameters, and obtain the upper limit on the
coupling parameter $\beta$ as
\begin{equation}
\beta < 1.11,~ 1.36,~ 1.53~ \,\, (2\sigma),
\end{equation}
and the present mass of neutrinos is also limited to 
\begin{equation}
\Omega_\nu h^2_{\rm{today}} < 0.0095,~ 0.0090,~ 0.0084~ \,\, (2\sigma),
\end{equation}
for models I, II and III, respectively.
\begin{widetext}
\end{widetext}
\begin{center}
\vskip1.0cm
\begin{figure}[t]
\vspace{0cm} \epsfxsize=5cm \centerline{
\rotatebox{0}{\includegraphics[width=0.3\textwidth]{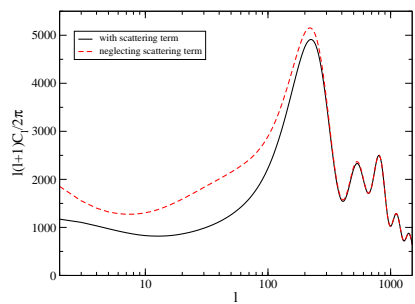}} }
\epsfxsize=5cm \caption{Differences between the CMB power spectra
with and without the scattering term in the geodesic equation of
neutrinos with the same cosmological parameters.
\label{fig:comparison-cmb-spectrum} }
\end{figure}
\end{center}
{\it \bf Results and discussions:}
Here we discuss two important points of this work and results of our analysis :
(a) the impact of the scattering term of the Boltzmann Equation, 
(b) the instability issue in our models, and 
(c) neutrino mass bounds in the interacting neutrino dark-energy models.

{\it \bf (a) Impact of the new scattering term:}
Recently, perturbation equations for the MaVaNs models were
nicely presented by Brookfield et al. \cite{Brookfield-b}, (see also
\cite{Zhao:2006zf}) which are necessary to compute CMB and LSS
spectra. A main difference here from their works is that we correctly
take into account the scattering term in the geodesic equation of
neutrinos, which was omitted there (see, however, \cite{Brookfield:2005bz}).
Because the term is proportional to
$\frac{\partial m}{\partial x}$ and first order quantity in
perturbation,  our results  and those of earlier works
\cite{Brookfield-b,Zhao:2006zf} remain the same in the background 
evolutions. However, as will be shown in the appendix, neglecting
this term violates the energy momentum conservation law at linear
level leading to the anomalously large ISW effect. Because the term
becomes important when neutrinos become massive, the late time ISW
is mainly affected through the interaction between dark energy and
neutrinos. Consequently, the differences show up at large angular
scales. In Fig. (\ref{fig:comparison-cmb-spectrum}), the differences
are shown with and without the scattering term. The early ISW can
also be affected by this term to some extent in some massive
neutrino models and the height of the first acoustic peak could be
changed. However, the position of the peaks stays almost unchanged
because the background expansion histories are the same.

\begin{figure}
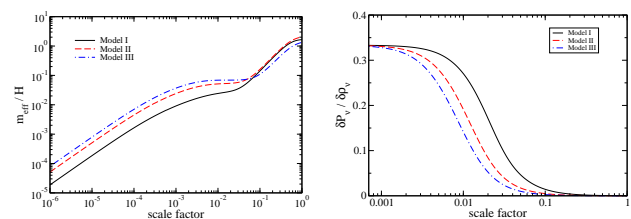

\begin{minipage}[m]{0.48\linewidth}
    \rotatebox{0}{\includegraphics[width=0.95\textwidth]{meffoH.eps}}
\end{minipage}
\begin{minipage}[m]{0.48\linewidth}
    \rotatebox{0}{\includegraphics[width=0.95\textwidth]{cs2_nu.eps}}
\end{minipage}
\caption{(Left panel): Typical evolution of the effective mass of the
 quintessence scalar field relative to theHubble scale, for all models
 considered in this paper. (Right panel): Typical evolution of the sound speed
 of neutrinos $c^s = \delta P_\nu/\delta\rho_\nu$ with the wavenumber
 $k=2.3\times 10^{-3}$ Mpc$^{-1}$, for models as
 indicated. The values stay 
 positive stating from $1/3$ (relativistic) and neutrinos are stable
 against the density fluctuation. 
}
\label{fig:m_eff}
\end{figure}

{\it \bf (b) Instability issue:} 
As shown in \cite{Afshordi:2005ym,Bean:2007ny}, some class of models with mass
varying neutrinos suffers from the adiabatic instability at the
first order perturbation level. This is caused by an additional force
on neutrinos mediated by the quintessence scalar field and occurs when
its effective mass is much larger than the hubble horizon scale, where
the effective mass is defined by $m_{\rm eff}^2=\frac{d^2 V_{\rm
eff}}{d\phi^2}$.  To
remedy this situation one should consider an appropriate quintessential
potential which has a mass comparable the horizon scale at present, and
the models considered in this paper are the case
\cite{Brookfield-b}. Interestingly, some authors have found that one can
construct viable MaVaNs models by choosing certain couplings and/or
quintessential potentials
\cite{Takahashi:2006jt,Kaplinghat:2006jk,Bjaelde:2007ki}. Some 
of these models even realises $m_{\rm eff} \gg H$. In  
Fig.(\ref{fig:m_eff}), masses of the scalar  
field relative to the horizon scale $m_{\rm eff}/H$ are plotted. We find that $m_{\rm eff}<H$ for almost all period and
the models are stable. We also dipict in Fig.(\ref{fig:m_eff}) the sound
speed of neutrinos defined by $c^2_s = \delta P_\nu/\delta\rho_\nu$ with
a wavenumber $k=2.3\times 10^{-3}$ Mpc$^{-1}$. 
\begin{figure}
\begin{minipage}[m]{0.48\linewidth}
    \rotatebox{0}{\includegraphics[width=1.0\textwidth]{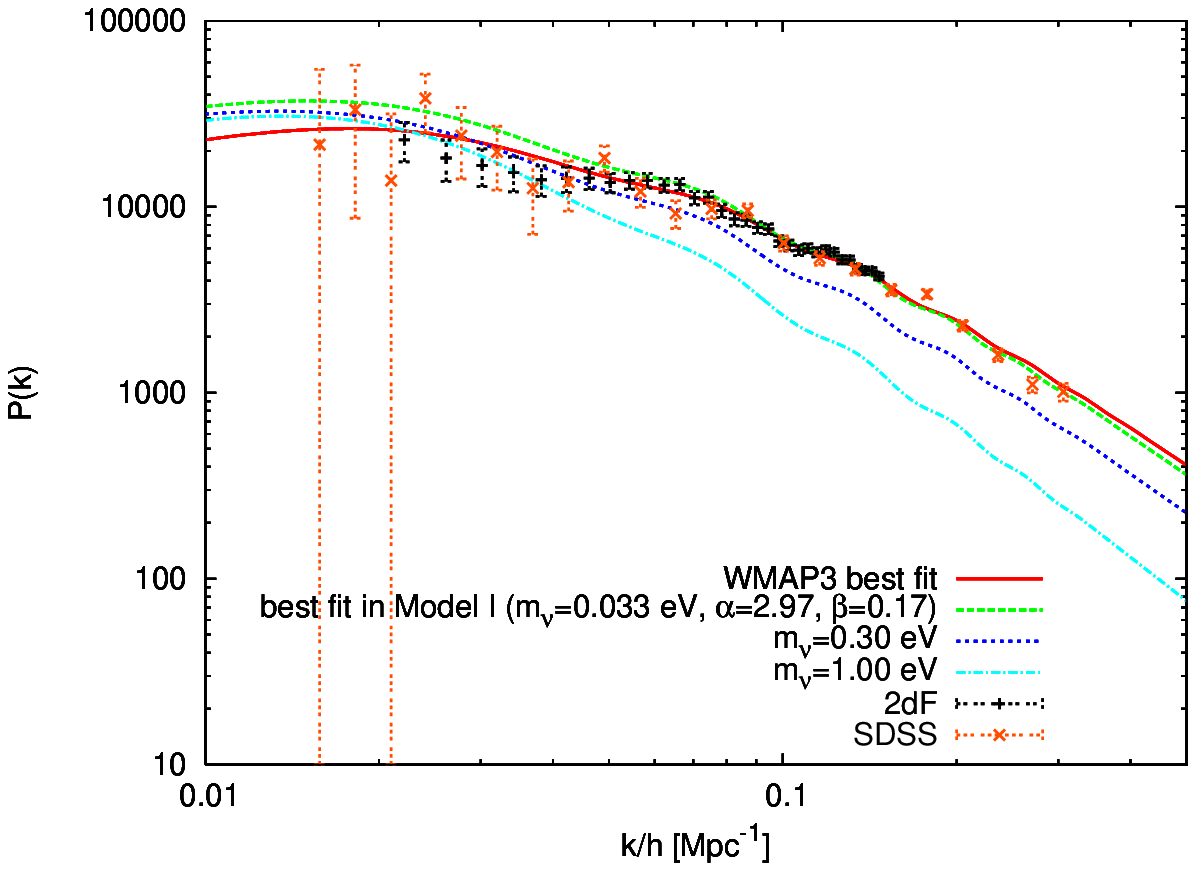}}
\end{minipage}
\begin{minipage}[m]{0.48\linewidth}
    \rotatebox{0}{\includegraphics[width=1.0\textwidth]{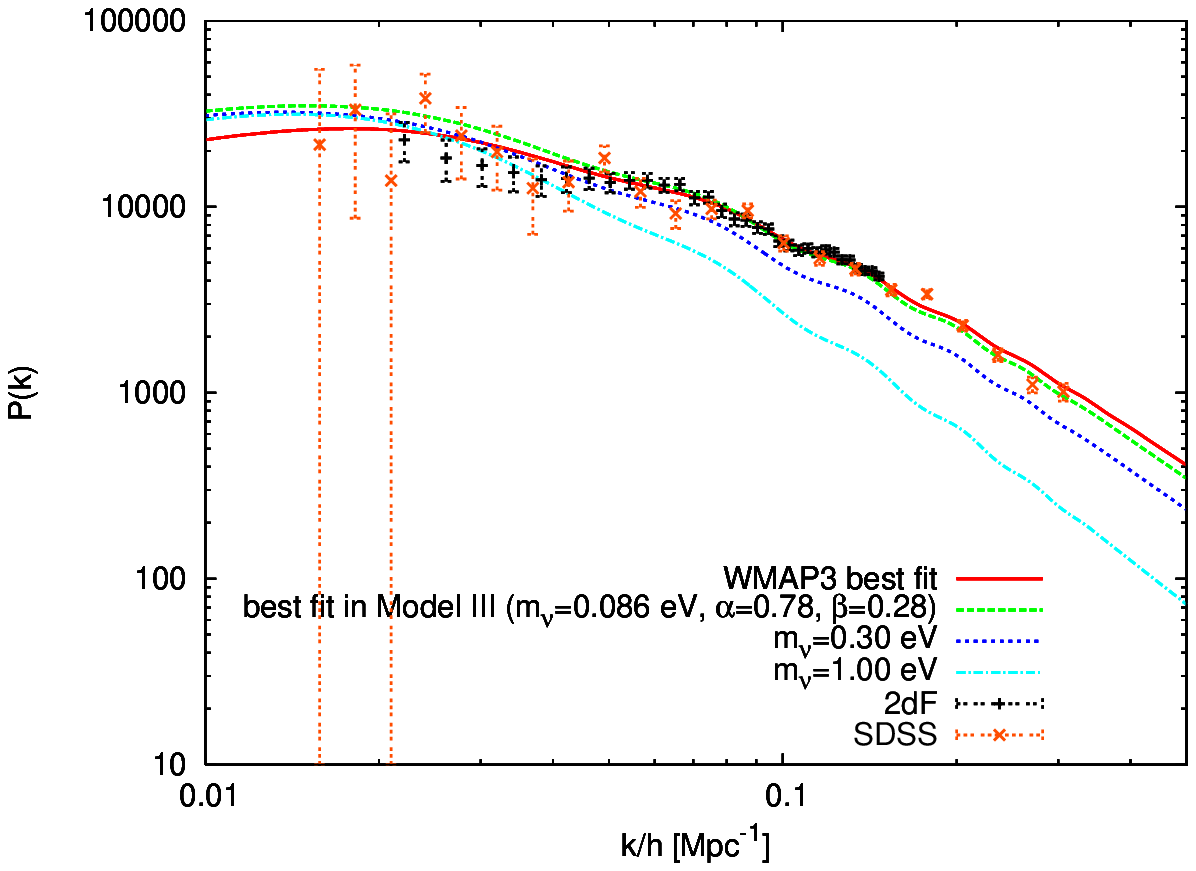}}
\end{minipage}
\caption{Examples of the total mass contributions in
the matter power spectrum in Model I (Left panel) and Model III
(Right panel). For both panels we plot the best fitting lines (green
dashed), lines with larger neutrino masses $M_\nu=0.3$ eV (blue
dotted) and $M_\nu=1.0$ eV (cyan dot-dashed) with the other
parameters fixed to the best fitting values. Note that while lines
with $M_\nu=0.3$ eV can fit to the data well by arranging the other
cosmological parameters, lines with $M_\nu=1.0$ eV can not.
 \label{fig:nu-mass-PS} }
\end{figure}

{\it \bf (c) Neutrino Mass Bounds:} 
When we apply the relation
between the total sum of the neutrino masses $M_{\nu}$ and their
contributions to the energy density of the universe:
$\Omega_{\nu}h^2=M_{\nu}/(93.14 eV)$, we obtain the constraint on
the total neutrino mass: $M_{\nu} < 0.87 eV (95 \% C.L.)$ in the
neutrino probe dark-energy model. The total neutrino mass
contributions in the power spectrum is shown in Fig
\ref{fig:nu-mass-PS}, where we can see the significant deviation
from observation data in the case of  large neutrino masses.

\acknowledgements
 We would like to thank 
L. Amendola, O. Seto, S. Carroll, and L. Schrempp  
for useful comments and discussions. 
K.I.'s work is supported by Grant-in-Aid for JSPS Fellows. 
Y.Y.K's work is
partially supported by Grants-in-Aid for NSC in Taiwan, Center
for High Energy Physics(CHEP)/KNU and APCTP in Korea.

\end{document}